\documentclass[%
superscriptaddress,
preprint,
 amsmath,amssymb,
]{revtex4-1}

\usepackage{graphicx}
\usepackage{dcolumn}
\usepackage{bm}

\begin{document}

\preprint{JAP}

\title{Localization of electronic states resulting from electronic topological transitions in the Mo$_{1-x}$Re$_x$ alloys: A photoemission study}


\author{L. S. Sharath Chandra} 
\email{lsschandra@rrcat.gov.in}%
\affiliation{Free Electron Laser Utilization Laboratory, Raja Ramanna Centre for Advanced Technology, Indore, Madhya Pradesh - 452 013, India}
\author{Shyam Sundar}
\address{Free Electron Laser Utilization Laboratory, Raja Ramanna Centre for Advanced Technology, Indore, Madhya Pradesh - 452 013, India}
\affiliation{Presently at Department of Physics, Simon Fraser University, Burnaby, British Columbia, Canada V5A 1S6}
\author{Soma Banik} 
\affiliation{Synchrotron Utilization Section, Raja Ramanna Centre for Advanced Technology, Indore, 452013, India}%
\affiliation{Homi Bhabha National Institute, Training School Complex, Anushakti Nagar, Mumbai 400 094, India}
\author{SK. Ramjan}
\address{Free Electron Laser Utilization Laboratory, Raja Ramanna Centre for Advanced Technology, Indore, Madhya Pradesh - 452 013, India}
\affiliation{Homi Bhabha National Institute, Training School Complex, Anushakti Nagar, Mumbai 400 094, India}
\author{M. K. Chattopadhyay}
\address{Free Electron Laser Utilization Laboratory, Raja Ramanna Centre for Advanced Technology, Indore, Madhya Pradesh - 452 013, India}
\affiliation{Homi Bhabha National Institute, Training School Complex, Anushakti Nagar, Mumbai 400 094, India}
\author{S. N. Jha}
\affiliation{BARC Beamline Section, Raja Ramanna Centre for Advanced Technology, Indore, Madhya Pradesh - 452 013, India.}
\affiliation{Atomic and Molecular Physics Division, Bhabha Atomic Research Centre, Mumbai 400085, India}
\affiliation{Homi Bhabha National Institute, Training School Complex, Anushakti Nagar, Mumbai 400 094, India}
\author{S. B. Roy}
\affiliation{UGC-DAE Consortium for Scientific Research, Khandwa Road, Indore, Madhya Pradesh- 452017}

\date{\today}

\begin{abstract}
We present the results of resonant photoemission spectroscopy experiments on the Mo$_{1-x}$Re$_{x}$ alloy compositions spanning over two electronic topological  transitions (ETT) at the critical concentrations $x_{C1}$ = 0.05 and $x_{C2}$ = 0.11.  The photoelectrons show an additional resonance ($R3$) in the constant initial state (CIS) spectra of the alloys along with two resonances ($R1$ and $R2$) which are similar to those observed in molybdenum. All the resonances show Fano-like line shapes. The asymmetry parameter $q$ of the resonances $R1$ and $R3$ of the alloys is observed to be large and negative. Our analysis suggests that the origin of large negative q is associated with phonon assisted inter band scattering between the Mo-like states and the narrow band that appeared due to the ETT.

\end{abstract}

\maketitle


\section{Introduction}
Excellent mechanical properties at elevated temperatures and room temperature workability of the Mo$_{1-x}$Re$_{x}$ alloys find widespread applications in medical fields, aerospace and defense industries and welding production \cite{war93, man03, mao08}. These alloys were also found to be promising materials for superconducting applications \cite{shy13, and89, vib14}. This is due to the occurrence of the electronic topological transitions (ETT) which improves the mechanical and superconducting properties when rhenium is added to molybdenum \cite{shy15, shy15a, ign80, dav70, smi76}. An ETT is a transition where pockets of Fermi surface appear or disappear when an external parameter such as composition, pressure, and/or magnetic field is varied \cite{lif60, vol17}. The ETT was theoretically predicted first by I.M. Lifshitz  \cite{lif60} for pure metals subjected to elastic strains. The first experimental evidence for the ETT was given by Brandt et. al., by analyzing the pressure induced changes in the superconducting properties of the Tl-Hg alloys \cite{bra66}.  In the Mo$_{1-x}$Re$_x$ alloys, the superconducting transition temperature ($T_C$) increases non-uniformly from 0.90~K for $x$ = 0 to about 12.6~K for the $x$ = 0.40 alloy \cite{ign80, shy15a} without a change in the crystal structure. The range of compositions where the sharp change in $T_C$ is observed is associated with two ETTs at the critical concentrations $x_{C1}$ = 0.05 and $x_{C2}$ = 0.11 \cite{oka13,ign07,vel86,gor91,sko94,ign02,sko98}. Earlier, we have shown that the ETTs and the superconducting properties are coupled in the Mo$_{1-x}$Re$_x$ alloys \cite{shy15a, shy16}. Our previous studies \cite{shy15, shy16} revealed that the appearance of Re 5$d$ like states at the Fermi level above $x > x_{C2}$ leads to multi-band superconductivity \cite{tar19}. We have also shown that the scattering of $s$ like electrons to Re 5$d$ like states by the soft phonon modes is responsible for the enhancement of $T_C$ for $x > x_{C2}$ \cite{shy15a}. The observation of the fact that the stress required to generate a fixed amount of strain $>$~3\% is minimum around 7 at.\% rhenium in molybdenum \cite{dav70} is due to the softening of phonons. The phonon softening improves the ductility of these alloys \cite{wad86}. Smith et al., have observed that the phonons soften along the N-H direction of the Brillouin zone when these alloys undergo ETT \cite{smi76}. This is the same location of the Brillouin zone where a pocket of the Fermi surface appears \cite{oka13} when more than 5 at.\% of rhenium is added to molybdenum. These features are associated with the changes in its electronic structure due to the ETT and the resulting localization of the electronic states for a small group of carriers that appear against the background of a continuous electronic spectrum \cite{ign07}. This localization of electronic states responsible for the ETT occurs due to the random potential introduced in the system when the composition is changed \cite{mot90, and58}. Localization of electronic states associated with the ETT also occurs when the pressure or magnetic field is a control parameter. This is due to the localization of electronic states at the band edges \cite{mot90}. However, the localization effects in the vicinity of the ETT is quite small and the detection requires very sensitive techniques \cite{ign07}. The helium ion channeling experiments on  Mo$_{1-x}$Re$_x$ alloys revealed that the effective mass of the electrons in the states of the new Fermi surfaces formed during the ETT are higher than the other electrons \cite{dik06}.  Ignat'eva and co-workers have also observed large  oscillations in the pressure dependence of $T_C$ and in certain temperature derivatives of the normal state thermoelectric power and resistivity against a specific background which is related to the ETT in the Mo$_{1-x}$Re$_x$ alloys \cite{ign07}. They argued that the localization of electrons filling the new states in the new Fermi surfaces appearing during the ETT cause the observed oscillations in the physical quantities \cite{ign07}. Here, we provide a direct experimental evidence of the localization of these electrons in the newly formed Fermi surfaces appearing during the ETT.

 The localized states against a background continuum give rise to Fano resonance in many of the observables \cite{fan61, mir10, mis15}. The photoelectrons from the valance band state with orbital angular momentum $l_v$ can show a Fano resonance when the photons of energies ($E_P$) corresponding to an inner core shell having an orbital angular momentum $l_i$ = $l_v$-1 are used for the photoemission spectroscopy (PES) measurement \cite{dav86, all92}. The interference between the electrons from the direct emission and those from the autoionization due to the super-Coster-Kronig transition can be explained by the Fano resonance \cite{dav86} as 

 \begin{equation}
I(E_P) = I_{nr}(E_P) + I_0(E_P )\frac{(\epsilon +q)^2}{1+\epsilon^2}.
 \end{equation} 

Here $I$($E_P$) is the photoemission intensity at a given binding energy of the valance band. The intensities $I_0$($E_P$) and $I_{nr}$($E_P$) correspond respectively to the transitions to the states of continuum which do and do not interact with the discrete autoionizing state. The $\epsilon$ = 2($E_P$-$E_R$)/$\Gamma$ is the reduced energy with  resonance energy $E_R$  and width $\Gamma$. The asymmetry parameter $q$ of the Fano resonance depends on the ratio of probabilities of transition to a discrete state and transition to the continuum as well as on the hybridization between the discrete and the continuum \cite{dav86, all92, mir10, mis15, som16, som17}.  When $|q| >>$ 1, the transition to the continuum is very weak and the line shape is determined only by the discrete state. The $|q|\approx$ 1 indicates a strong hybridization between the discrete and continuum states and $|q|$ = 0 indicates that the states belong to the continuum \cite{dav86, all92}. Therefore, the resonant PES technique (RPES) can be used to study the interaction of localized states with the continuum \cite{dav86, all92, som16, som17}. We therefore study the RPES of the bcc Mo$_{1-x}$Re$_x$ alloys around $x$ = $x_{Ci}$. 

Recently, we have shown that multi-band superconductivity manifests in the Mo$_{1-x}$Re$_x$ alloys with $x > x_{C2}$ \cite{shy15}. The $T_C$, the Sommerfeld coefficient of specific heat ($\gamma$) \cite{shy15a} and the elastic constant ($C_{11}$) \cite{sha18} are observed to change abruptly across $x_{C2}$. Our studies revealed that the origin of the above observations is related to the phonon assisted inter-band $s$-$d$ scattering between the band that newly appeared at the Fermi level ($E_f$) due to the ETT and the rest of the bands \cite{shy15a}. We have also shown from RPES studies on the alloys with $x >> x_{C2}$ that the newly appeared band has Re5$d$ like character \cite{shy16}. Further analysis by Evans and Dowben revealed stronger Mo-Re orbital hybridization for $x >> x_{C2}$ in comparison with the Mo-Mo or Re-Re bonding \cite{eva17}. 

In this article, we present the appearance of a distinct Fano like resonance in the constant initial state (CIS) intensities in the spectra of RPES measurements on the Mo$_{1-x}$Re$_x$ alloys for the density of states (DOS) at the binding energy ($E_B$) $\approx$ -2~eV below the $E_f$ when $x \geq x_{C1}$ = 0.05. Our analysis suggests that the observation of a large negative $q$ is associated with the localization of electron like states in the newly appeared Fermi pocket as well as with the phonon assisted inter-band $s$-$d$ interaction.      

\section{Experimental Details}
The arc melted polycrystalline samples of Mo$_{1-x}$Re$_{x}$ ($x$ = 0-0.15) alloys formed in the body centred cubic (bcc) phase (space group: Im$\bar{3}$m) \cite{shy15a}. The resonant photoemission measurements were performed at the Angle Resolved Photoelectron Spectroscopy beamline of Indus-1 Synchrotron, India. Base vacuum during resonant photoemission measurement was 3 $\times$ 10$^{-10}$~mbar. The samples were cleaned in situ by sputtering. The absence of carbon 1s peak at 284~eV and oxygen 1s peak at 531~eV was ensured before the measurements. The valence band photoemission spectra were recorded using Phoibos 150 electron energy analyser (SPECS) with a typical resolution of 135~meV in the range $E_P$ = 23~eV to 70~eV. In this energy range, the photoemission spectra is more bulk sensitive \cite{shy16}. Core levels were studied using X-ray photoemission spectroscopy (XPS) with Mg $K_\alpha$ source (XR 50, SPECS).

\section{Results and Discussions}

\begin{figure}[]
\includegraphics[width = 8.5cm,height = 7cm]{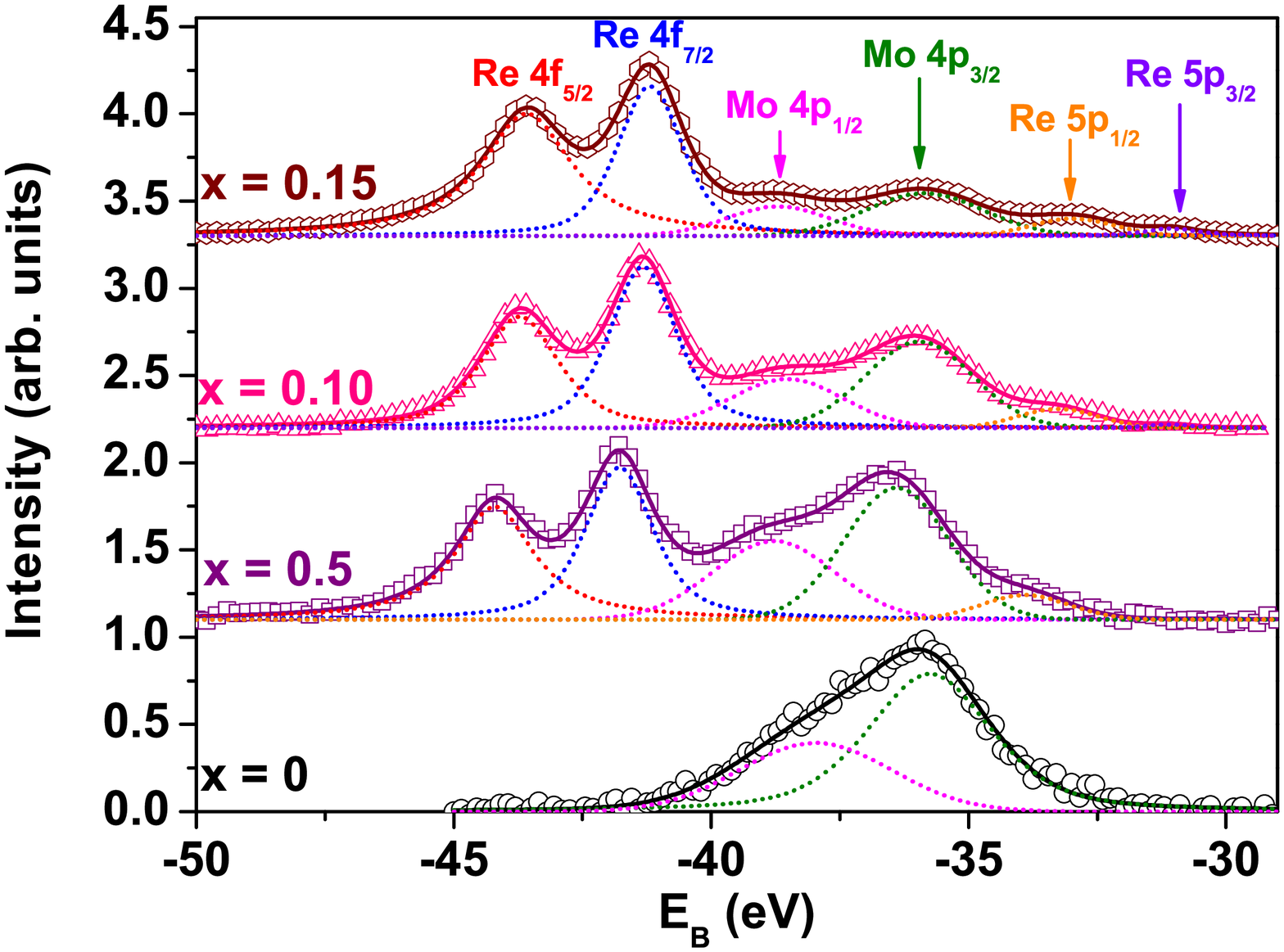}
\caption{\label{fig:epsart} X-ray photoelectron spectroscopy of the Mo$_{1-x}$Re$_{x}$ alloys corresponding to the Re5$p$, Mo4$p$ and Re4$f$ inner core shells having binding energies within the range -30 to -50~eV. The symbols represent the experimental spectra and the lines are the multi-peak fits used to extract the peak positions.}
\end{figure}

\begin{figure*}[]
	\includegraphics[width = 17cm,height = 17cm]{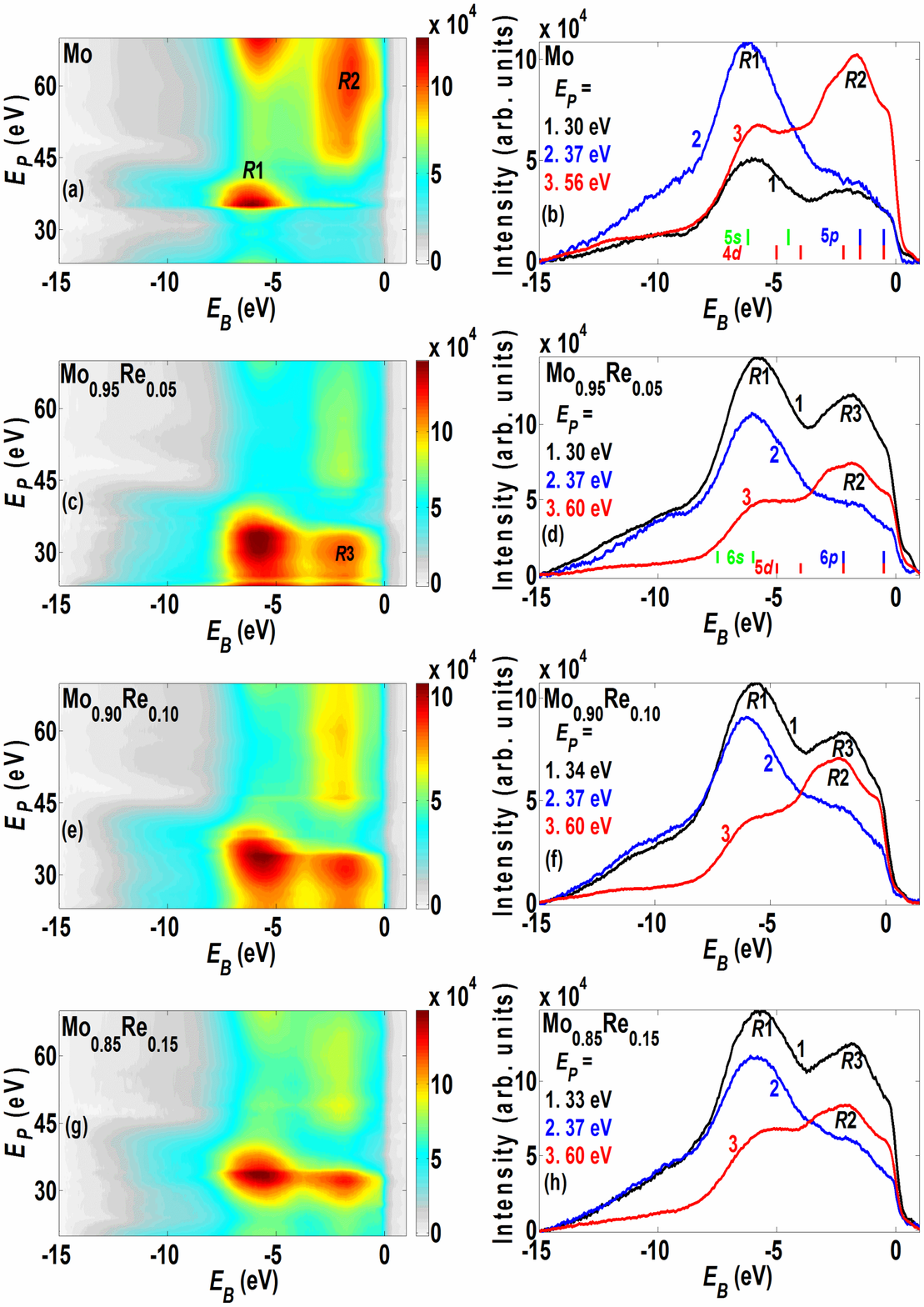}
	\caption{\label{fig:epsart} Photoelectron spectra of the valance band of Mo$_{1-x}$Re$_{x}$ at room temperature for (a,b) $x$ = 0, (c,d) $x$ = 0.05, (e,f) $x$ = 0.10 and (g,h) $x$ = 0.15. The images in the left are contour plots while the graphs in the right are photoemission spectra at a few selected $E_P$. Two resonances are seen in Mo while three resonances are observed in the alloys.}
\end{figure*}

Figure 1 shows the XPS spectra of the Mo$_{1-x}$Re$_{x}$ ($x$ = 0-0.15) alloys corresponding to the Re5$p$, Mo4$p$ and Re4$f$ inner core shells \cite{shy16, fuk80}. All these shells show spin-orbit (SO) splitting. The quantitative analysis is carried out using the XPSPEAK4.1 software. For elemental molybdenum, the positions of the Mo4$p_{3/2}$ and Mo4$p_{1/2}$  peaks are respectively at $E_B$ = -35.8~eV and -38~eV. The Re4$f$ inner core shell of the $x$ = 0.05 alloy shows two SO split peaks at $E_B$ = -41.8~eV and -44.2~eV respectively for  the Re4$f_{7/2}$ and Re4$f_{5/2}$ states. The $E_B$ of Re4$f$ core shell moves towards $E_f$ with increasing $x$. The features corresponding to Re5$p_{3/2}$ and Re5$p_{1/2}$ inner core shells \cite{fuk80} at $E_B$ = -31~eV and -33~eV are weakly visible for the alloys. As $x$ increases, the intensity of the Re4$f$ shell becomes predominant while that of the Mo4$p$ shell reduces and becomes feeble at $x$ = 0.15. Thus, the contributions of molybdenum and rhenium to the valance band of Mo$_{1-x}$Re$_{x}$ ($x$ = 0-0.15) alloys across the ETT can be studied through RPES in the range $E_P$ = 20-70~eV. In this range, the resonant enhancement of the valance band states of the Mo$_{1-x}$Re$_{x}$ alloys can be obtained from the interference of electronic wave functions from (a) direct photoemission and (b) Auger emissions for the following transitions \cite{shy16}:

(i) Mo 4$p$-5$s$ transition via 
\begin{eqnarray*}
\begin{aligned}
4p^64d^55s^1 + h\nu~&\rightarrow(a)~4p^64d^55s^0 + e^-\\ 
&\rightarrow(b)~4p^54d^55s^2~\rightarrow~4p^64d^55s^0 + e^-\\
\end{aligned}
\end{eqnarray*}
 (ii) Mo 4$p$-4$d$ transition via 
\begin{eqnarray*}
\begin{aligned}
4p^64d^55s^1 + h\nu~&\rightarrow(a)~4p^64d^45s^1 + e^-\\
&\rightarrow(b)~4p^54d^65s^1~\rightarrow~4p^64d^45s^1 + e^-\\
\end{aligned}
\end{eqnarray*}
and\\
(iii) Re $5p$-$5d$ transition via 
\begin{eqnarray*}
\begin{aligned}
5p^65d^56s^2 + h\nu~&\rightarrow(a)~ 5p^65d^46s^2 + e^-\\
&\rightarrow(b)~ 5p^55d^66s^2~\rightarrow~5p^65d^46s^2 + e^-\\
\end{aligned}
\end{eqnarray*}

Figure 2 shows the valance band photoemission spectra of the Mo$_{1-x}$Re$_{x}$ ($x$ = 0-0.15) alloys as a function of $E_P$. The Fermi energy ($E_f$) is taken as $E_B$ = 0. The valance band spectra of molybdenum (Fig. 2(a) and 2(b)) show two broad features at about $E_B$ $\approx$ -2~eV and $E_B$ $\approx$ -6~eV. As the photon energy is increased from  $E_P$ = 23~eV, the intensities of both the features are decreased. The intensity of the feature at $E_B$ $\approx$ -6~eV increases sharply and shows a resonance ($R1$) when the $E_P$ reaches the threshold energy (35~eV) for the Mo4$p$ inner core shell. The feature at $E_B$ $\approx$ -2~eV do not show appreciable resonance at $E_P \approx$ 35~eV. Above $E_P$ = 40~eV, the intensity of the feature at $E_B$ $\approx$ -2~eV increases slowly and shows a broad resonance ($R2$) around 45~eV.  The intensity of the feature at $E_B$ $\approx$ -6~eV again increases above 56~eV as $E_P$ approaches the Mo4$s$ threshold. These results are consistent with our previous studies on Mo \cite{shy16}. In Fig. 2(b) we have marked the positions ($E_B$(DFT)) of the Mo$4d$, Mo$5s$ and Mo$5p$ states where the allowed states in the valence band are expected from the density functional theory \cite{shy16}. From the comparison of  position of $E_B$ of the resonances with the $E_B$(DFT), we can conclude that the resonance $R1$ corresponds to the 4$p$ to 5$s$ transition \cite{shy16} while the delayed resonance $R2$  corresponds to the 4$p$ to 4$d$ transition. In comparison with the elemental molybdenum, substantial changes in the valance band photoemission of $x$ = 0.05 (Fig. 2(c) and 2(d)), 0.1(Fig. 2(e) and 2(f)) and 0.15 (Fig. 2(g) and 2(h)) alloys are observed in the $E_P$ range 23-38~eV. The  resonance $R1$ becomes broader with increasing $x$ (compare Fig.2 (a), (c) and (e)) and the valance band states around $E_B$ = -2~eV show additional resonance (marked as $R3$ in Fig. 2(c)) at about $E_P \approx$ 30~eV. The sharpness of the resonance $R3$ increases with increasing $x$, and hence, is related to the Re partial DOS. We have also marked the $E_B$(DFT) of the Re$5d$, Re$6s$ and Re$6p$ sub bands \cite{shy16} in the Fig. 2(d). As considerable part of the Re$5d$ sub-band is centred around $\approx$ -2~eV below $E_f$, the resonance $R3$ can be assigned to Re $5p$-$5d$ transition. For $E_P >$ 38~eV, the valance band spectra of the alloys are similar to that of molybdenum.  

\begin{figure}[]
\includegraphics[width = 8.5cm,height = 10cm]{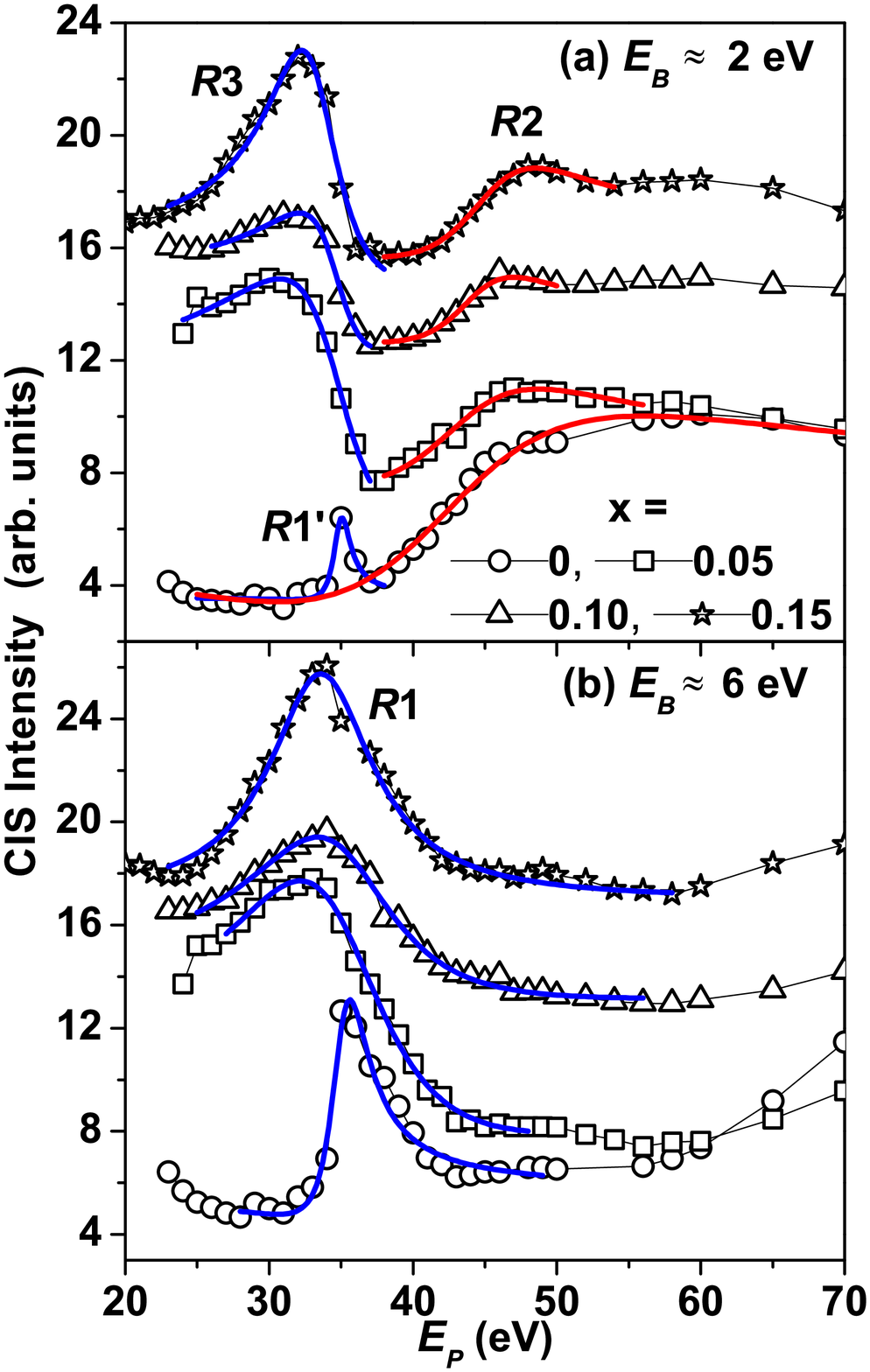}
\caption{\label{fig:epsart} Analysis of the constant initial state spectra at (a) $E_B \approx$ 2~eV and (b)  $E_B \approx$ 6~eV of the Mo$_{1-x}$Re$_{x}$ ($x$ = 0.05-0.15) alloys using eq.1. The symbols are the experimental data points and the lines are fits to the data. The fitting parameters are listed in table I. The $q$ for the resonance at $E_P \approx$ 34~eV in the alloys is large and negative. }
\end{figure}

Figure 3 shows the Fano line shape fitting of the constant initial state (CIS) intensities at (a) $E_B$$\approx$ -2~eV and (b) $E_B$$\approx$ -6~eV of valance band photoemission of the Mo$_{1-x}$Re$_{x}$ alloys. The parameters $E_R$, $\Gamma$ and $q$ of the fitting for ($x$ = 0, 0.05, 0.10 and 0.15) are presented in table I. The CIS plot of molybdenum for $E_B$$\approx$ -2~eV shows a weak peak ($R1'$) at $E_P \approx$ 34~eV (Fig.3(a)). The strength of this resonance increases with increasing $E_B$ and is maximum for $E_B$$\approx$ -6~eV ($R1$ in Fig. 3(b)). We found that molybdenum has only 5$s$ states at $E_B$$\approx$ -6~eV and the resonance $R1$ results from the 4$p$ to 5$s$ transition. The $q$ of $R1$ is about 3 over the entire valance band which indicates that the hybridization is weak between 4$d$ and 5$s$ states in the $E_B$ range -5~eV to $E_f$. The enhanced intensity of $R1$ at about $E_B$ = -6~eV indicates that most of the $5s$ states are present in this narrow energy range. The shift of the $R2$ resonance of the $4p$-$4d$ transition at $E_B \approx$ -2~eV from the Mo$4p$ threshold to $E_P \approx$ 45~eV indicates the presence of electron-electron correlations in the Mo$4d$ band \cite{dav86}. The $q$ of the $R2$ is $<$1 which indicates that the Mo4$d$ states have nearly free electron like character. In the alloys, the resonance $R2$ becomes sharper ($\Gamma$($x \neq$0)$<$ 0.5$\Gamma$(0)) with a $q \approx$ 1.2 due to the preferential Mo-Re bonding over the Mo-Mo or Re-Re bonding \cite{eva17}. The shape of the resonance $R1$ in the alloys is quite different from that of molybdenum. This indicates that the Re states contribute to $R1$ of the alloys. The $q$ of $R1$ in the alloys is negative and the $|q|$ increases sharply with increasing $x$. The value of $|q|$ for $x$ = 0.15 is tending to infinity indicating the localization of discrete Re states associated with the $R1$ resonance. 

\begin{table}[]
\caption{\label{tab:table1}The parameters corresponding to the fitting of the constant initial state spectra of the Mo$_{1-x}$Re$_{x}$ ($x$ = 0-0.15) alloys using eq.1. The fitting is shown in Fig. 3.}
\begin{ruledtabular}
\begin{tabular}{cccc}
$x$&$E_P$ (eV)&$\Gamma$ (eV)&$q$\\
\colrule
$R1$&&&\\
\colrule
0& 35.1$\pm$0.2&3.1$\pm$0.4&2.9$\pm$0.4\\
0.05& 34.9$\pm$0.4&13.5$\pm$0.5&-2.4$\pm$0.3\\
0.10&35$\pm$0.3&12.6$\pm$0.5&-3.8$\pm$0.5\\
0.15&35.6$\pm$0.2&8.8$\pm$0.5&$-\infty$\\
\colrule
$R2$&&&\\
\colrule
0& 41.8$\pm$1&25.3$\pm$1.2&0.89$\pm$0.1\\
0.05& 43.4$\pm$1&14.8$\pm$3.3&1.4$\pm$0.4\\
0.10&44.4$\pm$0.8&8.8$\pm$0.9&1.7$\pm$0.6\\
0.15&45.3$\pm$0.4&10.5$\pm$0.6&1.6$\pm$0.2\\
\colrule
$R3$&&&\\
\colrule
0.05& 34.8$\pm$0.4&8.8$\pm$1.6&-1.1$\pm$0.2\\
0.10&34.6$\pm$0.2&5.6$\pm$0.7&-1.1$\pm$0.1\\
0.15&33.3$\pm$0.2&6$\pm$0.7&-2.9$\pm$0.4\\
\end{tabular}
\end{ruledtabular}
\vskip -0.5 cm
\end{table}
The resonance $R3$ is observed only in the alloys and is quite different from the other resonances discussed above.  We see that the $E_R$ of $R3$ decreases with increasing $x$ and approaches the threshold of the Re$5p$ core shell for $x$ = 0.15. The values of $q$ of $R1$ and $R3$ of the alloys are large and negative. The large value of $q$ suggest that the bands responsible for the resonances are narrow and localized. The sign of the $q$ depends on the nature of interaction (mixing) between the discrete states and the continuum \cite{mis15, cha78}. The Fano line shapes with negative (positive) values of $q$ have been reported in the Raman lines originating from the discrete optical phonon mode and an electron (holes) continuum in $n$-Si ($p$-Si) \cite{cha78, sax17}. The change in the sign of $q$ between $p$-Si to $n$-Si arises from the difference in the nature of electron-phonon and hole-phonon interactions. The elemental molybdenum is a compensated metal with higher mobility of holes in comparison with the electrons \cite{cox72, cox73}. The addition of rhenium (with one extra electron in comparison with molybdenum) increases the number of electrons with an additional band crossing the Fermi level for $x > x_{C1}$ \cite{vel86, gor91, sko94, ign02, ign07, sko98, oka13}. We have shown that the scattering of electrons between this Re5$d$ like band and the rest of the bands is through the electron-phonon interaction (phonon assisted interband scattering (PAIS))\cite{shy15a}. Hence, the origin of negative $q$ observed for $R3$ in the alloys can be assigned to the phonon assisted $s-d$ interaction. We have also shown that the enhancement in the $T_C$ of the Mo$_{1-x}$Re$_{x}$ alloys is due to the enhancement of the electron phonon coupling constant corresponding to PAIS \cite{shy15a}. Therefore, the enhanced $T_C$ along with the two-gap superconductivity in the Mo$_{1-x}$Re$_{x}$ alloys are due to the localization of electrons in the new Fermi pockets formed due to the ETT.

\section{Conclusions}

The addition of rhenium to molybdenum is known to improve the ductility ("The Rhenium Effect"). It is observed that the stress required to produce a fixed amount of strain higher than 3\% is minimum around $x$ = 0.07 \cite{dav70}.  Smith et al. have observed that the phonons soften along the N-H direction of the Brillouin zone of the Mo$_{1-x}$Re$_{x}$ alloys when these alloys undergo ETT \cite{smi76}. This is the same location of the Brillouin zone where a pocket of the Fermi surface appears when $x >$ 0.05. Earlier, we have shown that the large enhancement of $T_C$ for $x >$ 0.05 is due to the changes in the electronic structure across the ETT \cite{shy15}. We have now shown that  the resonant photoemission technique may be effectively used to distinguish between the narrow localized states and the delocalized continuum states of the Mo$_{1-x}$Re$_{x}$ alloys resulting from the ETT at $x_{C1}$ = 0.05 and $x_{C2}$ = 0.11. The states that crosses the Fermi level due to the ETT show an additional resonance as compared to those observed for elemental molybdenum. The $q$ parameter of this resonance is large and negative indicating that these states are electron like and are localized. By comparing the present results with previous studies \cite{shy15, ign07}, the enhanced superconducting transition temperature and other functional properties of the Mo$_{1-x}$Re$_{x}$ alloys are linked to the interaction between these localized states with the rest of the delocalized states and the associated changes in the electron-phonon interaction.

\begin{acknowledgments}
The authors thank Babita Vinayak Salaskar for her help during experiments, Tapas Ganguli for his interest in this work, and Pankaj Sagdeo, IIT Indore for helpful discussion.
\end{acknowledgments}

\nocite{*}

\begin{thebibliography}{}
\bibitem{war93} See, e g., papers by J. Wardsworth and J. P. Wittenauer  and by R. L. Heenstand {\it Evolution of Refractory Metals and Alloys} eds. E N C Dalder, T Grobstein and C S Olson (Warrendale: The Minerals, Metals and Materials Society (1993) ).

\bibitem{man03} Mannheim R L and Garin J L 2003 {\it J. Mater. Process. Technol.} {\bf 143-144} 533 

\bibitem{mao08} Mao P, Han K and Xin Y 2008 {\it J. Alloys and Comp.} {\bf 464} 190

\bibitem{shy13} Shyam Sundar, L. S. Sharath Chandra, V. K. Sharma, M. K. Chattopadhyay, and S. B. Roy, AIP Conf. Proc. {\bf 1512} 1092 (2013). 

\bibitem{and89} A. Andreone, A. Barone, A. Di Chiara, F. Fontana, G. Mascolo, V. Palmieri, G. Peluso, G. Pepe, and U. Scotti Di Uccio, J. Supercond. {\bf 2} 493 (1989).

\bibitem{vib14} V. Singh, B. H. Schneider, S. J. Bosman, E. P. J. Merkx, and G. A. Steele, Appl. Phys. Lett. {\bf 105} 222601 (2014). 

\bibitem{shy15} Shyam Sundar, L. S. Sharath Chandra, M. K. Chattopadhyay and S. B. Roy, J. Phys. Condens. Mater {\bf 27}, 045701 (2015).

\bibitem{shy15a} Shyam Sundar, L. S. Sharath Chandra, M. K. Chattopadhyay S. K. Pandey, D. Venkateshwarlu, R. Rawat, V. Ganesan, and S. B. Roy, New J. Phys. {\bf 17}, 053003 (2015).

\bibitem{ign80} Ignat'eva T A and Cherevan' Yu A 1980 {\it Pis'ma Zh. Eksp. Teor. Fiz.} {\bf 31} 389 

\bibitem{dav70} Davidson D L and Brotzen F R 1970 {\it Acta Metallurgica} {\bf 18} 463

\bibitem{smi76} Smith H G, Wakabayashi  N and Mostoller M 1976 \textit{Proceedings of Second Rochester Conference on Superconductivity in d- and f - band metals} eds. Douglass D H (Prenum press, New York) p. 223

\bibitem{lif60} I. M. Lifshitz, J. Exptl. Theoret. Phys. {\bf 11}, 1130 (1960) [Zh. Eksp. Teor. Fiz. {\bf 38}, 1569 (1960)].

\bibitem{vol17} G. E. Volovik, Low Temp. Phys. {\bf 43} 47 (2017).

\bibitem{bra66} N. B. Brandt, N. I. Ginzburg, B. G. Lazarev, L. S. Lazareva, V. I. Makarov, and T. A. Ignat'eva, J. Exptl. Theoret. Phys. {\bf 22}, 61 (1966) [Zh. Eksp. Teor. Fiz. {\bf 49}, 85 (1965)]. 

\bibitem{ign07} T. A. Ignat'eva, Phys. Solid State {\bf 49}, 403 (2007) [Fiz. Tve. Tela {\bf 49}, 389 (2007)].

\bibitem{oka13} M. Okada, E. Rotenberg, S. D. Kevan, J. Schafer, B. Ujfalussy, G. M. Stocks, B. Genatempo, E. Bruno, and E. W. Plummer, New J. Phys. {\bf 15}, 093010 (2013). 

\bibitem{vel86} A. N. Velikodny, N. V. Zavaritskii, T. A. Ignat'eva, and A. A. Yurgens, Pis'ma Zh. Eksp. Teor. Fiz. {\bf 43}, 597 (1986).  
 
\bibitem{gor91} Y. N. Gornsoostyrev, M. I. Katsnelson, G. V. Peschanskikh, and A. V.  Trefilov, Phys. Stat. Sol. (b) {\bf 164}, 185 (2011).  

\bibitem{sko94} N. V. Skorodumova, S. I. Simak, Ya M. Blanter, and Yu Kh. Vekilov, Pis'ma Zh. Eksp. Teor. Fiz. {\bf 60}, 549 (1994).   

\bibitem{ign02} T. A. Ignat'eva, and A. N. Velikodny, Low Temp. Phys. {\bf 28}, 403 (2002).   

\bibitem{sko98} N. V. Skorodumova, S. I. Simak, I. A. Abrikosov, B. Johansson, and Yu. Kh. Vekilov, Phys. Rev. B {\bf 57}, 14673 (1998).   

\bibitem{shy16} Shyam Sundar, Soma Banik, L. S. Sharath Chandra, M. K. Chattopadhyay, Tapas Ganguli, G. S. Lodha, S. K. Pandey, D. M. Phase, and S. B. Roy, J. Phys. Condens. Mater {\bf 28}, 315502 (2016). 

\bibitem{tar19} V. Tarenkov, A. Dyachenko, V. Krivoruchko, A. Shapovalov and M. Belogolovskii, J. Supercond. Novel Mag. (2019). (DOI: 10.1007/s10948-019-05297-0).

\bibitem{wad86} J. Wadsworth, T. G. Nieh, and J.J. Stephens, Scripta Metallug.  {\bf 20}, 637 (1986). 

\bibitem{mot90} See. e. g. N. F. Mott, {\it Metal-Insulator Transitions} (2nd Ed., Taylor \& Francis, London (1990)).

\bibitem{and58} P. W. Anderson, Phys. Rev. {\bf 109}, 1492 (1958).

\bibitem{dik06} N. P. Dikiy, and T. A. Ignatyeva, Phys. Solid State {\bf 48}, 25 (2006).

\bibitem{fan61} U. Fano, Phys. Rev. {\bf 124}, 1866 (1961).

\bibitem{mir10} A. E. Miroshnichenko, S. Flach, Y. S. Kivshar, Rev. Mod. Phys. {\bf 82}, 2257 (2010).

\bibitem{mis15} O. V. Misochko, and M. V. Lebedeva, J. Exptl. Theoret. Phys. {\bf 120}, 651 (2015) [Zh. Eksp. Teor. Fiz. {\bf 147}, 750 (2015)].

\bibitem{dav86} L. C. Devis, J. Appl. Phys. {\bf 59}, R25 (1986).

\bibitem{all92} J. W. Allen, in {\it Ch6 of Synchrotron Radiation Research: Advances in surface and interface science} (Plenum Press, New York) {\bf 1}, pp. 253-326 (1992).

\bibitem{som16} D. Mondal, Soma Bnaik, C. Kamal, M. Nand, S. N. Jha, D. M. Phase, A. K. Sinha, A. Chakrabarti, A. Banerjee, and T. Ganguli, J. Alloy. Comp. {\bf 688}, 187 (2016).

\bibitem{som17} Soma Banik, P. K. Das, A. Bendounan, I. Vobornik, A. Arya, N. Beaulieu, J. Fujii, A. Thamizhavel, P. U. Sastry, A. K. Sinha, D. M. Phase, and S. K. Deb, Sci. Rep. {\bf 7}, 4120 (2017).

\bibitem{sha18} L. S. Sharath Chandra, Shyam Sundar, M. K. Chattopadhyay, and S. B. Roy, (unpublished (2018)). 

\bibitem{eva17} P. Evans, and P. A. Dowben, J. Phys. Condens. Mater {\bf 29}, 098001 (2017). 

\bibitem{fuk80} Y. Fukuda, F. Honda, and J. W. Rabalais, Surf. Sci. {\bf 93}, 338 (1980).

\bibitem{cha78} M. Chandrasekhar, J. B. Renucci, and M. Cardona, Phys. Rev. B {\bf 17}, 1623 (1978).

\bibitem{sax17} S. K. Saxena, P. Yogi, S. Mishra, H. M. Rai, V. Mishra, K. Warshi, S. Roy, P. Mondal, P. R. Sagdeo, and R. Kumar, Phys. Chem. Chem. Phys. {\bf19}, 3178895 (2017).

\bibitem{cox72} W. R. Cox, and F. R. Brotzen, J. Phys. Chem. Solids {\bf 33}, 2311 (1972).

\bibitem{cox73} W. R. Cox, D. J. Hayes, and F. R. Brotzen, Phys. Rev. B {\bf 7}, 3580 (1973).






\end{thebibliography}

\end{document}